\documentclass{ws-p8-50x6-00}

\begin{document}

\title{Non--conventional statistical effects in relativistic 
heavy-ions collisions}

\author{W.M. Alberico}
\address{Dipartimento di Fisica, Universit\`a di Torino, Italy\\
INFN - Sezione di Torino, Italy}

\author{A. Lavagno$^1$ and P. Quarati$^2$}
\address{Dipartimento di Fisica, Politecnico di Torino, Italy\\
$^1$INFN, Sezione di Torino, Italy\\
$^2$INFN, Sezione di Cagliari, Italy}  

\maketitle

\abstracts{
We show that non--conventional statistical effects 
(due to the presence of long range forces, memory effects,  
correlations and fluctuations)  
can be very relevant in the interpretation of the experimental observables in 
relativistic heavy-ions collisions.  
Transverse mass spectrum, transverse momentum fluctuations and 
rapidity spectra are analysed in the framework of the non-extensive 
statistical mechanics. }

\section{Introduction and motivations}

Most of the theoretical analyses related to observables in 
relativistic heavy-ion 
collisions involve (implicitly or explicitly) the validity 
of the standard Boltzmann-Gibbs statistical mechanics. In particular, if the 
thermal equilibrium is achieved, the Maxwell-Boltzmann (MB) distribution 
(Fermi-Dirac or Bose-Einstein distribution if quantum statistical effects 
are not negligible) is assumed to hold. 
When the system approaches equilibrium, 
the phase--space distribution should be derived 
as a stationary state of the dynamical kinetic evolution equation.
It is well known that in the absence of non--Markovian memory 
effects, long--range interactions and local 
correlations, the MB distribution is obtained as a steady state solution of 
the kinetic Boltzmann equation. 
However, it is a rather common opinion that, 
because of the extreme conditions of 
density and temperature in ultrarelativistic heavy ion collisions, 
memory effects and long--range color interactions give rise to the presence 
of non--Markovian processes in the kinetic equation affecting the 
thermalization process toward equilibrium as well as the standard 
equilibrium distribution \cite{biro,ropke,roberts,albe}. 

The aim of the present contribution is to explore, 
from a phenomenological 
point of view, the relevance of the above mentioned statistical effects that 
can influence the dynamical evolution of the generated fireball toward the 
freeze-out stage and, as a consequence, the physical observables.

\section{Generalized non--extensive statistics}

A quite interesting generalization of the conventional Boltzmann--Gibbs 
statistics has been recently proposed by Tsallis \cite{tsa} and proves to 
be able to overcome the  shortcomings  of the conventional statistical 
mechanics in many physical problems, where the presence of 
long--range interactions, long--range microscopic memory, or fractal 
space--time constraints hinders the usual statistical assumptions.
 
The Tsallis generalized thermostatistics  is based upon the following 
generalization of the entropy \cite{tsa} 
\begin{equation}
S_q=\frac{1}{q-1}\, \sum_{i=1}^W p_i \, (1-p_i^{q-1}) \;,
\label{tsaen}
\end{equation}
where $p_i$ is the probability of a given microstate among $W$ different ones 
and $q$ is a fixed real parameter.
The new entropy has the usual properties of positivity, equiprobability, 
concavity and irreversibility, preserves the whole mathematical 
structure of thermodynamics and
reduces to the conventional Boltzmann--Gibbs entropy $S=-\sum_i p_i \log p_i$
in the limit $q\rightarrow 1$.

The single particle distribution function is obtained  through the 
usual procedure of maximizing the Tsallis entropy  under the 
constraints of keeping constant 
the average internal energy and the average number of particles. 
For a dilute gas of particles and/or for $q\approx 1$ values, the 
average occupational number can be written in a simple analytical 
form 
\begin{equation}
\langle n_i\rangle_q=\frac{1}{[1+(q-1)\beta (E_i-\mu)]^{1/(q-1)}\pm 1} \;,
\label{distri}
\end{equation}
where the  $+$ sign is for fermions, the $-$ for bosons and $\beta=1/T$.
In the limit 
$q\rightarrow 1$ (extensive statistics), one recovers the conventional 
Fermi--Dirac and Bose--Einstein distribution. 
Under the same conditions, but in the classical limit, 
one has the following  generalized Maxwell--Boltzmann 
distribution \cite{tsa}:

\begin{equation}
\langle n_i\rangle_q=[1+(q-1)\beta (E_i-\mu)]^{1/(1-q)} \;.
\label{distribz}
\end{equation}
When the entropic $q$ parameter is smaller than $1$, the distributions 
`(\ref{distri}) and (\ref{distribz}) have a natural high energy cut--off: 
$E_i\le 1/[\beta (1-q)]+\mu$,  which implies that the energy tail is 
depleted; when $q$ is greater than $1$, the cut--off is absent and the energy 
tail of the particle distribution (for fermions and bosons) is enhanced. 
Hence the nonextensive statistics entails a sensible difference of the 
particle distribution shape in the high energy region with respect 
to the standard statistics.
This property plays an important r\^ole in the interpretation of the 
physical observables, as it will be shown in the following.

\section{Transverse mass spectrum and momentum fluctuations}

Let us consider the 
transverse momentum distribution of particles produced, e.g., in relativistic 
heavy ion collisions: it depends on the phase-space distribution 
and usually an exponential shape is employed to fit the experimental data. 
This shape is obtained by assuming a purely thermal source with a MB  
distribution. High energy deviations from the exponential shape are taken 
into account by introducing a dynamical effect due to collective transverse 
flow, also called blue-shift. 

Let us consider a different  point of view and argue that
if long tail time memory and long--range interactions are present,  
the MB distribution must be replaced by
the generalized distribution (\ref{distribz}).
Limiting ourselves to consider here only small deviations from 
standard statistics ($q-1\approx 0$); then 
at  first order in 
$(q-1)$ the transverse mass spectrum can be written as 
\begin{equation}
\frac{dN}{m_\perp dm_\perp}=C \; m_\perp \left\{ K_1\left ( z \right )+ 
\frac{(q-1)}{8} z^2 \; \left [3 \; K_1 (z)+K_3 (z) \right ]  \right\} \;,
\label{mpt}
\end{equation}
where $K_i$ are the modified Bessel function at the $i$-order. 

The above equation is able to reproduce very well the transverse momentum 
distribution of hadrons produced in S+S collisions 
(NA35 data \cite{na35}) providing we take $q=1.038$ 
\cite{albe2}. 
Furthermore, it is easy to see that at first 
order in $(q-1)$ from Eq.(\ref{mpt}), 
the generalized slope parameter takes the following form:
%
\begin{equation}
T_q=T+ (q-1) \, m_\perp \; .
\label{qtslope}
\end{equation}
Hence nonextensive statistics predicts, in a purely thermal source, 
a generalized $q$--blue shift factor 
at high $m_\perp$; moreover  this shift factor is 
not constant but increases (if $q>1$) with $m_\perp=\sqrt{m^2+p_\perp^2}$, 
where $m$ is the mass of the detected particle. Such a behavior has been 
observed in the experimental NA44 results\cite{na44}. 

Another observable very sensitive 
to non--conventional statistical effects   
are the transverse momentum fluctuations. 
In fact, in the framework of non--extensive statistics, 
the particle fluctuation 
$\langle\Delta n^2\rangle_q=\langle n^2\rangle_q-\langle n \rangle^2_q$ 
is deformed, with respect to the standard expression, as follows
\begin{equation}
\langle\Delta n^2\rangle_q\equiv
\frac{1}{\beta}\frac{\partial\langle n\rangle_q}{\partial\mu}=
\frac{\langle n\rangle_q }{1+(q-1)\beta (E-\mu)}\, (1\mp \langle n\rangle_q)\;,
\label{fluc}
\end{equation}
where $E$ is the relativistic energy $E=\sqrt{m^2 + p^2}$. 

The fluctuations of an ideal gas of fermions (bosons), expressed 
by Eq.(\ref{fluc}), are still suppressed (enhanced) by the factor 
$1\mp\langle n\rangle_q$ (as in the standard case) but 
this effect is modulated by the 
factor $[1+(q-1) \beta (E-\mu)]^{-1}$.
Therefore the fluctuations turn out to be increased 
for $q<1$ and are decreased for $q>1$.
Very good agreement with the experimental NA49 analysis \cite{na49b} 
is obtained by taking $q=1.038$ \cite{albe}, notably 
the same value used in transverse momentum spectra.

\section{Anomalous diffusion in rapidity spectra}

An important observable in relativistic heavy-ion collisions is 
the rapidity distribution of the detected particles. 
In particular, there is experimental and theoretical evidence 
that the broad rapidity 
distribution of net proton yield ($p-\overline{p}$) in central 
heavy-ion collisions at SPS energies could be a signal of  
non-equilibrium properties of the system. 
We want to show now that the broad rapidity shape can be well 
reproduced in the framework of a non-linear relativistic 
Fokker-Planck dynamics which incorporates non-extensive 
statistics and anomalous diffusion. 

A class of anomalous diffusions are currently described through the 
non-linear Fokker-Planck equation (NLFPE)
\begin{equation}
\frac{\partial}{\partial t}[f(y,t)]^{\mu}=\frac{\partial}{\partial y} 
\left [ J(y) [f(y,t)]^\mu+D \frac{\partial}{\partial y}[f(y,t)]^\nu
\right ] \; ,
\label{nlfpe}
\end{equation}
where $D$ and $J$ are the diffusion and drift coefficients, respectively.
Tsallis and Bukman \cite{tsallis} have shown that, for 
linear drift, the time dependent solution of the above equation is a 
Tsallis-like distribution with $q=1+\mu-\nu$. The norm of the distribution is 
conserved at all times only if $\mu=1$, therefore we will 
limit the discussion to the case $\nu=2-q$.

Imposing the validity of the Einstein 
relation for Brownian particles, we can generalize to the relativistic case 
the standard expressions of diffusion and drift coefficients as follows 
\begin{equation}
D=\alpha \, T\; , \ \ \ \  \ J(y)=\alpha \, m_\perp \sinh(y)
\equiv  \alpha \, p_\parallel \; ,
\label{coeffi}
\end{equation}
where $p_\parallel$ is the longitudinal momentum and $\alpha$ 
is a common constant. 
It is easy to see that the above coefficients give 
us the Boltzmann stationary distribution in the 
linear Fokker-Planck equation ($q=\nu=1$) 
(such a result cannot be obtained if one assumes a linear drift 
coefficient as in Ref.\cite{wol})
while  the stationary solution of the NLFPE (\ref{nlfpe}) 
with $\nu=2-q$ is a Tsallis-like 
distribution with the relativistic energy $E=m_\perp \cosh(y)$: 

\begin{equation}
f_q(y,m_\perp)=
\Big \{1-(1-q)\,\beta\, m_\perp \cosh(y)\Big\}^{1/(1-q)} \; .
\end{equation}

Basic assumption of our analysis is that the rapidity distribution is 
not appreciably influenced by transverse dynamics, which is considered 
in thermal equilibrium. This hypothesis is well 
confirmed by the experimental data\cite{na44,na49} 
and adopted in many theoretical works\cite{braun}. 
Therefore, the time dependent rapidity distribution 
can be obtained, first, by means of numerical integration of 
Eq.(\ref{nlfpe}) with initial $\delta$-function condition depending 
on the value of the experimental projectile rapidities and, second, by 
integrating such a result over the transverse mass 
$m_\perp$ (or transverse momentum) as follows
\begin{equation}
\frac{dN}{dy}(y,t)=c\, \int_m^\infty \!\! m^2_\perp \, \cosh(y) 
\, f_q(y,m_\perp,t) \, dm_\perp \; ,
\label{raint}
\end{equation}
where $c$ is the normalization constant fixed 
by the total number of the particles. 
The calculated rapidity spectra will ultimately depend on the nonextensive 
parameter $q$ only, since there exists only one 
``interaction time'' $\tau=\alpha t$ which reproduces the experimental 
distribution \cite{lava}. 

\begin{figure}[ht]
\vspace{-1.5cm}
\epsfxsize=30pc 
\epsfbox{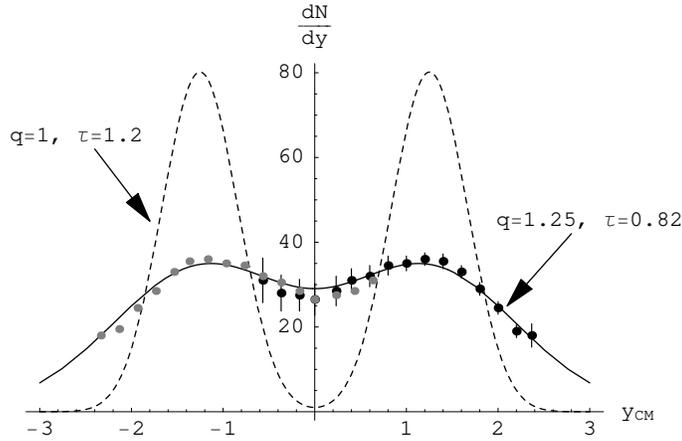} 
\vspace{-11cm}
\caption{Rapidity spectra for net proton production ($p-\overline{p}$) in 
central Pb+Pb collisions at 158A GeV/c (grey circles are data reflected about 
$y_{cm}=0$). Full line corresponds 
to our results by using a non-linear evolution equation ($q=1.25$), 
dashed line corresponds to the linear case ($q=1$).}
\end{figure}

In Fig.1 we show the calculated rapidity spectra of net proton 
compared with the experimental NA49 data from central  
Pb+Pb collisions at 158 GeV/c \cite{na49}.
The obtained spectra are normalized 
to 164 protons and the beam rapidity is fixed to $y_{\rm cm}=2.9$ (in the 
c.m. frame) \cite{na49}. 
The full line corresponds to the NLFPE solution (\ref{raint}) at 
$\tau=0.82$ and $q=1.25$; the dashed line corresponds to the solution 
of the linear case ($q=1$) at $\tau=1.2$. 
Only in the non-linear case ($q\ne 1$) there  
exists a (finite) time for which 
the obtained rapidity spectra well reproduces the broad experimental shape. 
A value of $q\ne 1$ implies anomalous superdiffusion in the rapidity space, 
i.e., $[y(t)-y_M(t)]^2$ scales like $t^\alpha$ with $\alpha>1$ \cite{tsallis}.

\section{Conclusions}

The nonextensive statistics appears suitable to evaluate physical observables
recently measured in heavy ion collision experiments. The physical motivation 
for the validity of a non-conventional statistical behavior can be related
to the presence of memory effects and long range interactions at 
the early stage of the collisions, 
even if a microscopic justification of these effects is still lacking. 
A rigorous determination of the conditions that produce a 
nonextensive statistical regime should be based on microscopic calculations 
relative to the parton plasma originated during the high energy collisions.
Non--perturbative QCD effects in the proximity of 
hadronic deconfinement could  
play a crucial role in the determination of the quantum kinetic evolution of 
the system toward the equilibrium \cite{roberts}.


\begin{thebibliography}{99}

\bibitem{biro}
T.S. Bir\'o and C. Greiner, \Journal{\PRL}{79}{3138}{1997}.
\bibitem{ropke}
S. Schmidt et al, \Journal{\PRD}{59}{94005}{1999}.
\bibitem{roberts}
D.V. Vinnik et al., {\em Eur. Phys. J.} C {\bf 22}, 341 (2001).
\bibitem{albe}
W.M. Alberico, A. Lavagno, P. Quarati, 
{\em Eur. Phys. J.} C {\bf 12}, 499 (2000).
\bibitem{tsa}
C. Tsallis, {\em J. Stat. Phys.} {\bf 52}, 479 (1988). \\ See also 
http://tsallis.cat.cbpf.br/biblio.htm for a regularly updated bibliography on 
the subject
\bibitem{na35} 
T. Alber et al. (NA35 Collab.), {\em Eur. Phys. J.} C {\bf 2}, 643 (1998). 
\bibitem{albe2}
W.M. Alberico, A. Lavagno, P. Quarati, 
{\em Nucl. Phys.} A {\bf 680}, 94 (2001).
\bibitem{na44}
I.G. Bearden et al. (NA44 Collab.), \Journal{\PRL}{78}{2080}{1997}.
\bibitem{na49b} 
A. Appelsh\"auser et al. (NA49 Collab.), \Journal{\PLB}{467}{21}{1999}.
\bibitem{tsallis}
C. Tsallis, D.J. Bukman, {\em Phys. Rev.} E {\bf54}, R2197 (1996). 
\bibitem{wol}
G. Wolschin, {\em Eur. Phys. J.} A {\bf 5}, 85 (1999).
\bibitem{na49}
H. Appelsh\"auser et al. (NA49 Collab.), \Journal{\PRL}{82}{2471}{1999}.
\bibitem{braun}
P. Braun-Munzinger et al., \Journal{\PLB}{344}{43}{1995}; 
\Journal{\PLB}{356}{1}{1996}.
\bibitem{lava}
A. Lavagno, {\em Physica} A {\bf 305}, 238 (2002).

\end{thebibliography}
\end{document}